\documentclass[11pt]{article}

\usepackage{amsfonts,amsthm,amsmath,amssymb,bbm,mathrsfs,verbatim,paralist}
\usepackage{epsfig,color}
\usepackage{graphicx}

\numberwithin{equation}{section}

\theoremstyle{plain}
\newtheorem{thm}{Theorem}[section]

\newtheorem{conj}[thm]{Conjecture}
\theoremstyle{definition}

\theoremstyle{remark}

\newcommand{\cconj}[1]{Conjecture~\ref{#1}}

\newenvironment{proofof*}[1]{\medskip\noindent
  \textit{Proof of #1:} }{}

\newcommand{\ie}{i.e.~}

\newcommand{\be}{\begin{equation}}
\newcommand{\ee}{\end{equation}}
\newcommand{\bea}{\begin{eqnarray}}
\newcommand{\eea}{\end{eqnarray}}
\newcommand{\nn}{\nonumber}

\def\rw{\rightarrow}

\newcommand\NN %\font\smallcal=cmsy8 \pagestyle{myheadings}
{{\mathbb N}}

\newcommand{\cE}{\mathcal{E}}
\newcommand{\cF}{\mathcal{F}}

\newcommand{\cS}{\mathcal{S}}

\def\a{\alpha}
\def\b{\beta}

\def\p{\rho}

\makeatletter
\def\underbracket{\@ifnextchar [
  {\@underbracket} {\@underbracket [\@bracketheight]}}
\def\@underbracket[#1]{\@ifnextchar [
  {\@under@bracket[#1]} {\@under@bracket[#1][0.2em]}}
\def\@under@bracket[#1][#2]#3{%\message {Underbracket: #1,#2,#3}
  \mathop {\vtop {\m@th \ialign {##
           \crcr $\hfil \displaystyle {#3}\hfil$
           \crcr \noalign {\kern 3\p@ \nointerlineskip }\upbracketfill 
{#1}{#2}
           \crcr \noalign {\kern 3\p@ }
           \crcr }}}\limits}
\def\upbracketfill#1#2{$\m@th \setbox \z@ \hbox {$\braceld$}
                 \edef\@bracketheight{0.5pt}%\the\ht\z@}
                 \upbracketend{#1}{#2}
                 \leaders \vrule \@height #1 \@depth \z@ \hfill
                 \leaders \vrule \@height #1 \@depth \z@ \hfill
                 \upbracketend{#1}{#2}$}
\def\upbracketend#1#2{\vrule height #2 width #1\relax}

\def\overbracket{\@ifnextchar [
{\@overbracket} {\@overbracket [\@bracketheight]}}
\def\@overbracket[#1]{\@ifnextchar [
{\@over@bracket[#1]} {\@over@bracket[#1][0.2em]}}
\def\@over@bracket[#1][#2]#3{%\message {Overbracket: #1,#2,#3}
\mathop {\vbox {\m@th \ialign {##
 \crcr \noalign {\kern 3\p@ }
 \downbracketfill {#1}{#2}
 \crcr \noalign {\kern 3\p@ \nointerlineskip }
 \crcr  $\hfil \displaystyle {#3}$%\hfil$
 \crcr
 } }}\limits}
\def\downbracketfill#1#2{$\m@th
 \setbox \z@ \vbox {$\braceld$}
 \edef\@bracketheight{0.5pt}%\the\ht\z@}
 \downbracketend{#1}{#2}
 \leaders \vrule \@height #1 \@depth \z@ \hfill
 \leaders \vrule \@height #1 \@depth \z@ \hfill
\downbracketend{#1}{#2}$}
\def\downbracketend#1#2{\vrule depth #2 width #1\relax}
\makeatother

\title{Phase diagram of a generalized ABC model on the interval \footnote{Dedicated to Cyril Domb, from whom we have learned so much.}}
\author{{ J. Barton$^{1}$, J. L. Lebowitz${}^{1,2}$,
and E. R. Speer${}^2$}\\ \\
{\small $^1$ Department of Physics, Rutgers University,}\\
{\small Piscataway, NJ 08854 USA}\\ 
{\small $^2$ Department of Mathematics, Rutgers University,}\\
{\small Piscataway, NJ 08854-8019 USA}}

\begin{document}

\maketitle

\begin{abstract}
We study the equilibrium phase diagram of a generalized ABC model on an interval of the one-dimensional lattice: each site $i=1,\ldots,N$ is occupied by a particle of type $\a=A,B,C,$ with the average density of each particle species $N_\a/N=r_\a$ fixed. These particles interact via a mean field non-reflection-symmetric pair interaction. The interaction need not be invariant under cyclic permutation of the particle species as in the standard ABC model studied earlier. We prove in some cases and conjecture in others that the scaled infinite system $N\rw\infty$, $i/N\rw x\in[0,1]$ has a unique density profile $\p_\a(x)$ except for some special values of the $r_\a$ for which the system undergoes a second order phase transition from a uniform to a nonuniform periodic profile at a critical temperature $T_c=3\sqrt{r_A r_B r_C}/2\pi$.
\end{abstract}

\noindent
{\bf Keywords:} generalized ABC model, external fields, phase diagram, scaling limit

\section{Introduction} \label{introduction}

The standard ABC model on an interval was considered in \cite{ACLMMS}. It is an equilibrium system on a 1D lattice of $N$ sites with closed boundary conditions.  Each site is occupied by one of three types of particles, denoted by $A$, $B$, and $C$.  The particles interact via a cyclic mean field type pair potential which is however not spatially reflection symmetric.  In this paper we generalize this model by introducing an additional interaction in which each of the particle types moves in a separate background potential that depends linearly on position. This breaks the cyclic symmetry in the standard ABC model. The equilibrium state of the standard ABC model may also be obtained as the steady state for certain nearest neighbor exchange dynamics \cite{ACLMMS,EKKM1,EKKM2}, and the generalized model considered here may be obtained by a modification of the exchange rates; see Appendix~\ref{microsystem} for details.

To define the model we introduce the occupation variables $\eta_\a(i)$, with $\eta_\a(i)=1\;(0)$ if site $i$ is (is not) occupied by a particle of type $\a$. As each site is occupied by exactly one particle,
\be \label{sumrho}
\sum_{\a}{\eta_\a(i)}=1.
\ee
The energy of a configuration $\underline{\eta}$ is defined to be
\be \label{microenergy}
E(\underline{\eta})=\frac{1}{N}\sum_{\a}{ \sum_{i=1}^N{ \left(\sum_{j=1}^N{ \Theta(j-i)\eta_\a(i)\eta_{\a+2}(j) } + i\,\xi_\a\,\eta_\a(i)\right) } },
\ee
Here $\a+1$ corresponds to the species following $\a$ in the ABC cyclic order, $\Theta(j-i)=1\;(0)$ for $j>i\;(j\leq i)$, and the $\xi_\a$  may be thought of as constant background electric fields, with $\xi_\a$ acting on particles of type $\a$.  We will consider the model in the canonical ensemble, with specified particle numbers of each type:
\be \label{canonical}
\sum_{i=1}^N{\eta_\a(i)}=N_\a,\qquad\sum_{\a}{N_\a}=N.
\ee
The canonical Gibbs measure for this system is then given by
\be \label{measure}
\mu_\b(\underline{\eta};{N_\a})=\frac{1}{Z}e^{-\b E(\underline{\eta})},
\ee
with $Z$ the usual canonical partition function. We will assume $N_\a>0$ for all $\a$ throughout; if one of the particle species is absent, the model simply reduces to the weakly asymmetric simple exclusion process (WASEP) \cite{SS,BE}.

Note that if one adds the same constant to each of the $\xi_\a$ then the energy \eqref{microenergy} is only changed by an overall constant.  We may therefore set $\sum_\a{\xi_\a}=0$, without loss of generality. We will refer to the case where $\xi_A=\xi_B=\xi_C=0$ as the ``standard'' ABC model. This is the model considered in \cite{ACLMMS}.

The energy \eqref{microenergy} may also be written in a different form, in which the contribution of the external fields is expressed through a modified mean field interaction. Using \eqref{sumrho} and \eqref{canonical} we have
\be \label{microidentity}
\begin{array}{l}
\displaystyle \sum_{i=1}^N{i\,\eta_\a(i)} = -\sum_{i=1}^N{ \sum_{j=1}^N{ \Theta(j-i)\eta_\a(i)\left(\eta_{\a+1}(j)+\eta_{\a+2}(j)\right) } }\\
\displaystyle \hskip90pt +\;N_\a\left(N-\frac{N_\a}{2}+\frac12\right).
\end{array}
\ee
Substituting \eqref{microidentity} in \eqref{microenergy} and rearranging sums, we obtain
\be \label{microenergy2}
\begin{array}{l}
\displaystyle E(\underline{\eta}) = \frac{1}{N}\sum_{\a}{ \Bigg( \sum_{i=1}^N{ \sum_{j=1}^N{  \Theta(j-i)\,3\,v_{\a+1}\,\eta_\a(i)\,\eta_{\a+2}(j) } } }\\
\displaystyle \hskip90pt + \; \xi_\a \frac{N_\a}{2}\left(N+N_{\a+2}-N_{\a+1}+1\right) \Bigg),
\end{array}
\ee
where the $v_\a$ are given by
\be \label{valpha}
v_\a=\frac{1}{3}\bigg(1+\xi_{\a+1}-\xi_{\a+2}\bigg),
\ee
so that
\be \label{sumvalpha}
\sum_\a{v_\a}=1.
\ee
The term $\sum_\a{\xi_\a\,N_\a\left(N+N_{\a+2}-N_{\a+1}+1\right)/2}$ is independent of the configuration, and may be ignored in the canonical ensemble with the $N_\a$ fixed.

\begin{figure}
\centerline{\includegraphics[scale=.5]{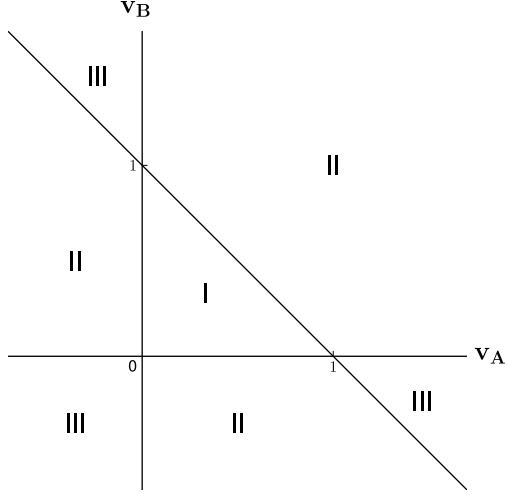}}
\caption{Different regions of the parameter space of the $v_\a$, plotted in the $v_A$-$v_B$ plane.}
\label{fig:regions}
\end{figure}

It will be convenient to consider the fundamental parameters of the model, in addition to the particle numbers $N_\a$ of each species, to be the $v_\a$ rather than the fields $\xi_\a$, as the $v_\a$ are more directly related to the physical behavior of the model.  We divide the space of these parameters into three regions; see Figure~\ref{fig:regions}, plotted in terms of $v_A$ and $v_B$, as these fix $v_C$ by \eqref{sumvalpha}.  In region I, $v_\a>0$ for all $\a$.  In regions II and III, $v_\a<0$ for one or two values of $\a$, respectively. In the standard model $v_A=v_B=v_C=1/3$.

We see from \eqref{microenergy2} that each of the $v_\a$ determines the energetically preferred alignment of the two other particle species $\a\pm1$. Effectively there is a contribution of $3\,v_\a/N$ to the energy every time any pair of particles of species $\a+1$ and $\a+2$ are not cyclically aligned, that is, whenever a particle of type $\a+2$ precedes one of type $\a+1$. If for example $v_A>0$, a pair of $B$ and $C$ particles will have a lower energy arranged as $B\cdots C$ than as $C\cdots B$. If $v_A<0$ the preferred arrangement is reversed, and the configuration $C\cdots B$ will have a lower energy than the usual cyclic ordering $B\cdots C$. This determines the ground states of the system, when $\b\rw\infty$, as described in Appendix~\ref{microsystem}.

\medskip
\noindent{\bf\large Background}
\medskip

The standard ABC model was introduced by Evans et al \cite{EKKM1,EKKM2} and in the form we use by Clincy et al \cite{CDE}. This model was originally considered on the ring, \ie with the boundary conditions periodic rather than closed, by specifying a dynamics consisting of asymmetric nearest neighbor exchanges between particles of different species. The stationary state of this dynamical system on the ring is generally not an equilibrium one. Its properties have been studied extensively in \cite{EKKM2} and in \cite{FF,BDLW,BdSGJL,LCM,BCP}. In the special case that the number of particles of each species is the same the stationary state of the model defined on the ring is a canonical Gibbs measure with the energy given by \eqref{microenergy2}, with $v_\a=1/3$ for all $\a$. The stationary state of the dynamical model defined on the interval, or equivalently on a ring with exchanges across one bond blocked, is always the canonical Gibbs measure regardless of the number of particles of each species and for equal $N_\a$ is identical to that on the ring.

The equilibrium properties of the standard ABC model on the interval were obtained exactly in \cite{ACLMMS}, see also \cite{BLS}. In particular it was shown there that (in the thermodynamic limit) the system has a unique state (density profiles) whenever the average densities $N_\a/N=r_\a$ are not all equal to $1/3$, \ie $r_\a\neq v_\a$ for some $\a$. When $r_\a=1/3$ the system undergoes a second order transition at $\b=\b_c=2\pi\sqrt{3}$ from a uniform density profile to a periodic profile \cite{BDLW}. There are thus for $\b>\b_c$ a continuum of phases (density profiles) specified by a rotation angle $\theta$. The results we derive here for the case when the $v_\a$ are not all equal to $1/3$ are more restricted. They suggest however that the phase diagram for general $v_\a$ is qualitatively similar to that of the standard ABC model, with a phase transition only at $r_\a=v_\a$.

The outline of the rest of the paper is as follows. In Section~\ref{scalinglimit} we discuss general properties of the macroscopic system in the scaling limit. Section~\ref{solutions} describes basic properties of the solutions of the Euler-Lagrange equations. Special cases are considered in Sections~\ref{rvequal} and \ref{specialcases} for different values of the parameters. In Appendix~\ref{microsystem} we discuss properties of the microscopic model specified by \eqref{microenergy}. Appendix~\ref{lvfamily} discusses the connection between the generalized ABC model and the Lotka-Volterra family of ODE systems, and in Appendix~\ref{excludedn} we discuss some restrictions on solutions of the Euler-Lagrange equations for the case $r_\a\neq v_\a$ for some $\a$, with the $v_\a$ in region I.

\section{Scaling limit} \label{scalinglimit}

The main goal of this paper is to study the phase diagram of the equilibrium system with energy \eqref{microenergy} when $N$ becomes macroscopic. For this we consider the scaling limit in which
\be \label{limit}
N\rw\infty,\qquad N_\a/N\rw r_\a,\qquad i/N\rw x\in[0,1],
\ee
so that $r_\a$ is the average density of the particle species $\a$. In this limit the state of the system is described by density profiles $\p_\a(x)$, $\a=A,B,C$, where $\p_\a(x)$ represents the density of particles of type $\a$ at a position $x$. These density profiles satisfy the constraints
\be \label{constraints}
0\leq\p_\a(x)\leq1,\qquad \int_0^1{\! \hbox{d}x\;\p_\a(x) }=r_\a,\qquad \sum_\a{\p_\a(x)}=1.
\ee

Let $\cS(\{\p\})$ and $\cE(\{\p\})$ be the entropy per site and energy per site in the scaling limit. As in \cite{ACLMMS}, see also \cite{CDE,BCP}, these are given by
\bea
\cS(\{\p\}) & = & -\sum_\a{ \int_0^1{\! \hbox{d}x\; \p_\a(x)\log{\p_\a(x)} } },\label{macroentropy}\\
\cE(\{\p\}) & = & \sum_\a{ \int_0^1{\! \hbox{d}x\! \int_0^1{\! \hbox{d}y\; \Theta(y-x)\p_\a(x)\p_{\a+2}(y) } } }\nn\\
            &   & \hskip40pt +\sum_\a{ \int_0^1{\! \hbox{d}x\; x\,\xi_\a\,\p_\a(x) } } + \rm{constant},\label{Econstant}\\
            & = & \sum_\a{ \int_0^1{\! \hbox{d}x\! \int_0^1{\! \hbox{d}y\; \Theta(y-x)\,3\,v_{\a+1}\,\p_\a(x)\p_{\a+2}(y) } } },\label{macroenergy}
\eea
where, as in the microscopic case, we have used the constraints \eqref{constraints} to rewrite the energy in a form analogous to \eqref{microenergy2}. The constant term in \eqref{Econstant}, equal to $-\sum_\a{\xi_\a r_\a(1+r_{\a+2}-r_{\a+1})/2}$, is independent of the profiles $\p_\a(x)$ and has been added in to simplify the expression \eqref{macroenergy}. The free energy (multiplied by $\b$) $\cF(\{\p\})$ associated to the canonical ensemble measure \eqref{measure}, is then given by
\be \label{energy}
\begin{array}{l}
\displaystyle \cF(\{\p\}) = \b\cE(\{\p\})-\cS(\{\p\})\\
\displaystyle \hskip38pt  = 3\b\sum_\a{ \int_0^1{\! \hbox{d}x\! \int_0^1{\! \hbox{d}y\; \Theta(y-x)v_{\a+1}\p_\a(x)\p_{\a+2}(y) } } }\\
\displaystyle \hskip80pt  + \sum_\a{ \int_0^1{\! \hbox{d}x\; \p_\a(x)\log{\p_\a(x)} } }.
\end{array}
\ee
%

%
%\be \label{energy2}
%\begin{array}{l}
%\displaystyle \cF(\{\p\}) = \b\sum_\a{ \int_0^1{\! \hbox{d}x\! \int_0^1{\! \hbox{d}y\; \Theta(y-x)\,3\,v_{\a+1}\,\p_\a(x)\p_{\a+2}(y) } } }\\
%\displaystyle \hskip80pt + \sum_\a{ \int_0^1{\! \hbox{d}x\; \p_\a(x)\log{\p_\a(x)} } },
%\end{array}
%\ee
%

The functional $\cF(\{\p\})$ is, up to an additive constant, the large deviation functional giving probabilities in the $N\rw\infty$ limit \cite{ACLMMS,CDE,BCP}, that is, the probability of the profile $\p_\a(x)$ is proportional to $\exp{\left[-N\left(\cF(\{\p\})-\min_\p{\cF(\{\p\})}\right)\right]}$. The typical equilibrium density profiles for the macroscopic model are thus those that minimize the free energy. There will be a coexistence of phases when the minimizer is not unique. To study this we have to consider solutions of the Euler-Lagrange equations (ELE) associated to the stationary points of \eqref{energy}.

Let $\cF_\a(x)=\delta\cF/\delta\p_\a(x)$ be the variational derivative of $\cF$ taken as though all of the $\p_\a$ are independent. Applying the constraints \eqref{constraints} leads to the ELE
\be \label{ELE0}
\cF_A-\cF_C=\hbox{constant},\qquad\cF_B-\cF_C=\hbox{constant},
\ee
where the $\cF_\a$ from \eqref{energy} are
\be \label{Falpha}
\begin{array}{l}
\displaystyle \cF_\a(x)=1+\log{\p_\a(x)}+3\b v_{\a+1}r_{\a+2}\\
\displaystyle \hskip80pt -3\b\int_0^x{\! \hbox{d}y\left(v_{\a+1}\p_{\a+2}(y)-v_{\a+2}\p_{\a+1}(y)\right) }.
\end{array}
\ee
It is easy to show that these $\cF_\a$ satisfy
\be
\sum_\a{\p_\a(x)\frac{\partial \cF_\a}{\partial x}}(x)=0,
\ee
which with \eqref{ELE0} implies that $\cF_\a(x)$ is constant for all $\a$. Then from \eqref{Falpha} the ELE are given by
\be \label{ELEint}
\p_\a(x)=\p_\a(0)\exp{\left(3\b\int_0^x{\! \text{d}y \left(v_{\a+1}\p_{\a+2}(y)-v_{\a+2}\p_{\a+1}(y)\right) } \right)},
\ee
which may be evaluated at $x=1$ to yield the boundary conditions
\be \label{BC}
\p_\a(1)=\p_\a(0)\,e^{3\b\left(v_{\a+1}r_{\a+2}-v_{\a+2}r_{\a+1}\right)}.
\ee
Equivalently, one may also write the ELE in differential form as
\be \label{ELE}
\frac{\hbox{d}\p_\a}{\hbox{d}x}=3\b\p_\a\left(v_{\a+1}\p_{\a+2}-v_{\a+2}\p_{\a+1}\right).
\ee
The ELE are to be solved subject to \eqref{constraints}. Solutions of \eqref{ELE} are stationary solutions of the hydrodynamic equations associated to the microscopic evolution of the model under the dynamics discussed in Appendix~\ref{microsystem} \cite{CDE,BCP}; these hydrodynamic equations have the form
\be
\frac{\partial \p_\a}{\partial t}=\frac{\partial}{\partial x}\left(\p_\a \frac{\partial \cF_\a}{\partial x} \right).
\ee

Examining the ELE \eqref{ELE} one finds that there exist two constants of the motion,
\be \label{sum}
\sum_\a{\p_\a(x)}=1,
\ee
and
\be \label{K}
K=\prod_\a{\p_\a^{v_\a}(x)}.
\ee
To study solutions of the ELE we eliminate $\p_C$ using \eqref{sum}, so that the solutions give trajectories in the $\p_A$-$\p_B$ plane, and more specifically in the triangle
\be \label{triangle}
\p_A,\p_B\ge0, \qquad \p_A+\p_B\le1.
\ee
Any such trajectory lies within a level set of $K$; as we will see below, when the $v_\a$ lie in region~I such a level set is either the single point $\p_\a=v_\a$ or a simple closed curve encircling this point.  When the $v_\a$ lie in regions II and III the level curve is an open curve joining two vertices of the triangle.  A solution of the ELE satisfying the boundary conditions \eqref{BC}, or equivalently the constraint $\int_0^1\p_\a(x)\,dx=r_\a$, is obtained by choosing first a value of $K$ and then a portion of the trajectory labeled by $K$ which is traversed in time one.  See Figures~\ref{fig:curves1} and \ref{fig:curves2} for typical level sets and trajectories.

In the special case where $r_\a=v_\a$ the macroscopic free energy is rotation invariant (see Appendix~\ref{microsystem} for details on the analogous result for the microscopic system).  To verify this we consider rotated profiles
\be
\tilde{\p}_\a(x)=\left\{\begin{gathered}
                        \p_\a(x-z)\;\;\;\;\;\;\;\;\;\;\;\;\hbox{if }x\geq z \hfill\\
                        \p_\a(x+(1-z))\;\;\;\hbox{if }x\leq z \hfill\\
                       \end{gathered} \;.
\right.
\ee
The entropy is clearly unchanged by the rotation, $\cS(\{\tilde{\p}\})=\cS(\{\p\})$, while the difference in energy is
\bea
\cE(\{\tilde{\p}\})-\cE(\{\p\}) & = & 3\sum_\a{ \int_{1-z}^{1}{\!\!\!\! \hbox{d}x\! \int_{0}^{1-z}{\!\!\!\!\!\!\! \hbox{d}y \; v_{\a+1}\left(\p_\a(x)\p_{\a+2}(y)-\p_\a(y)\p_{\a+2}(x)\right) } } }\nn\\
                               & = & 3\sum_\a{ \int_{1-z}^{1}{\!\!\!\! \hbox{d}x \; v_{\a+1}\left(\p_\a(x)\,r_{\a+2}-r_\a\,\p_{\a+2}(x)\right) } }=0,
\eea
where we have used $r_\a=v_\a$ and $\int_0^1{\!\hbox{d}x\;\p_\a(x)}=r_\a$. As we shall see this case, which generalizes the $r_\a=1/3$ case in the standard ABC model, plays a special role in the phase diagram.

It was proven in \cite[Section~10]{ACLMMS} for the standard ABC model that solutions of the ELE always exist, and that the minimizer of the free energy must be given by one of the solutions. The same result holds for the generalized ABC model, with no modification of the proof required. The fundamental question is then whether or not there is a unique minimizer.  This question was answered completely for the standard ABC model in \cite{ACLMMS}, and we believe that the direct generalization of the behavior in that case should hold for the general model. We state this here as a conjecture, although it is partially established in \cite{ACLMMS} and in the remainder of this paper; the statement  depends upon a critical temperature $T_c=\b_c^{-1}$, where
\be \label{betac}
\b_c=\frac{2\pi}{3\sqrt{v_A v_B v_C}}.
\ee

\begin{conj} \label{main} {\bf Solutions of the ELE.} (a) If the $v_\a$ lie in region~I and $r_\a=v_\a$ for $\a=A,B,C$ then there exist (i)~the constant solution $\p_\a(x)=v_\a$, $0\le x\le1$, (ii)~for $\beta>n\,\beta_c$, $n=1,2,\ldots$, a unique solution corresponding to a trajectory which traverses one of the level sets of $K$ exactly $n$ times, which we refer to as a type $n$ solution, and (iii)~no other solutions.  The minimizer of the free energy is, for $\beta\le\beta_c$, the (unique) constant solution and, for $\beta>\beta_c$, the type $1$ solution. At $\b_c$ there is a second order phase transition from the homogeneous phase to the phase segregated, heterogeneous phase.

\smallskip\noindent
(b) For values of the $v_\a$ and $r_\a$ other than those discussed in (a) there exists for every $\beta$ a unique solution minimizing the free energy.  \end{conj}

Statement (a.i) here is a trivial observation, and the existence portion of (a.ii) follows immediately from the existence of a minimizer and the fact that $\b>\b_c$ the uniform solution is unstable (Section~\ref{linearstability}); we give an independent argument in Section~\ref{rvequal}. Beyond this, as we will discuss below, we can prove all or part of this conjecture for some special values of the parameters besides the standard case, $v_\a=1/3$ for all $\a$, for which the conjecture has been proven in full. In particular we can prove uniqueness of the solution of the ELE when $\b<4\pi/3$. This follows from the result in \cite{ACLMMS} that for such $\b$ the standard ABC model free energy functional is globally convex on the space of density functions satisfying \eqref{constraints}. As the addition of external fields only adds terms that are linear in the particle densities to the standard ABC free energy (see \eqref{macroenergy}), the second variation of \eqref{energy} with respect to the density functions is the same as that of the standard model. Thus for $\b<4\pi/3$ the free energy for the generalized ABC model with external fields is also globally convex, implying that there is a unique solution of the ELE, which must be the minimizer of the free energy. For $r_\a=v_\a$ this is just the constant solution $\p_\a(x)=v_\a$, $0\leq x\leq1$. For other values of the $r_\a$ there is for $\b<4\pi/3$ a unique segment of a unique $K=K(\b;\underline{r})$ trajectory which minimizes $\cF$.

For $\b\geq 4\pi/3$ and $r_\a=v_\a$ (so that the $v_\a$ are in region I), we have the following additional results:
\begin{enumerate}[(i)]\itemsep1pt
\item \cconj{main}(a) is proven when, for some $\a$, $v_\a=1/2$ and $v_{\a+1}=v_{\a+2}=1/4$, and checked numerically in other cases..
\item The constant solution is linearly stable for $\b<\b_c$ and unstable for $\b>\b_c$.
\item Nonconstant minimizers are always of type $1$.
\item For small enough $K$, or equivalently, for large enough $\b$, the type $1$ solution is unique.
\end{enumerate}
When $r_\a\neq v_\a$ for some $\a$ we prove uniqueness for the cases when one of the $v_\a$ is zero and the other two have opposite signs, and when one of the $v_\a$ is one.

In the next section we describe general properties of the solutions of the ELE. The case $r_\a=v_\a$ (with the $v_\a$ in region I) is discussed in detail in Section~\ref{rvequal}, while the special cases when $r_\a\neq v_\a$ are considered in Section~\ref{specialcases}.

\section{General properties of solutions of the ELE} \label{solutions}

\begin{figure}
\centerline{\includegraphics[scale=.5]{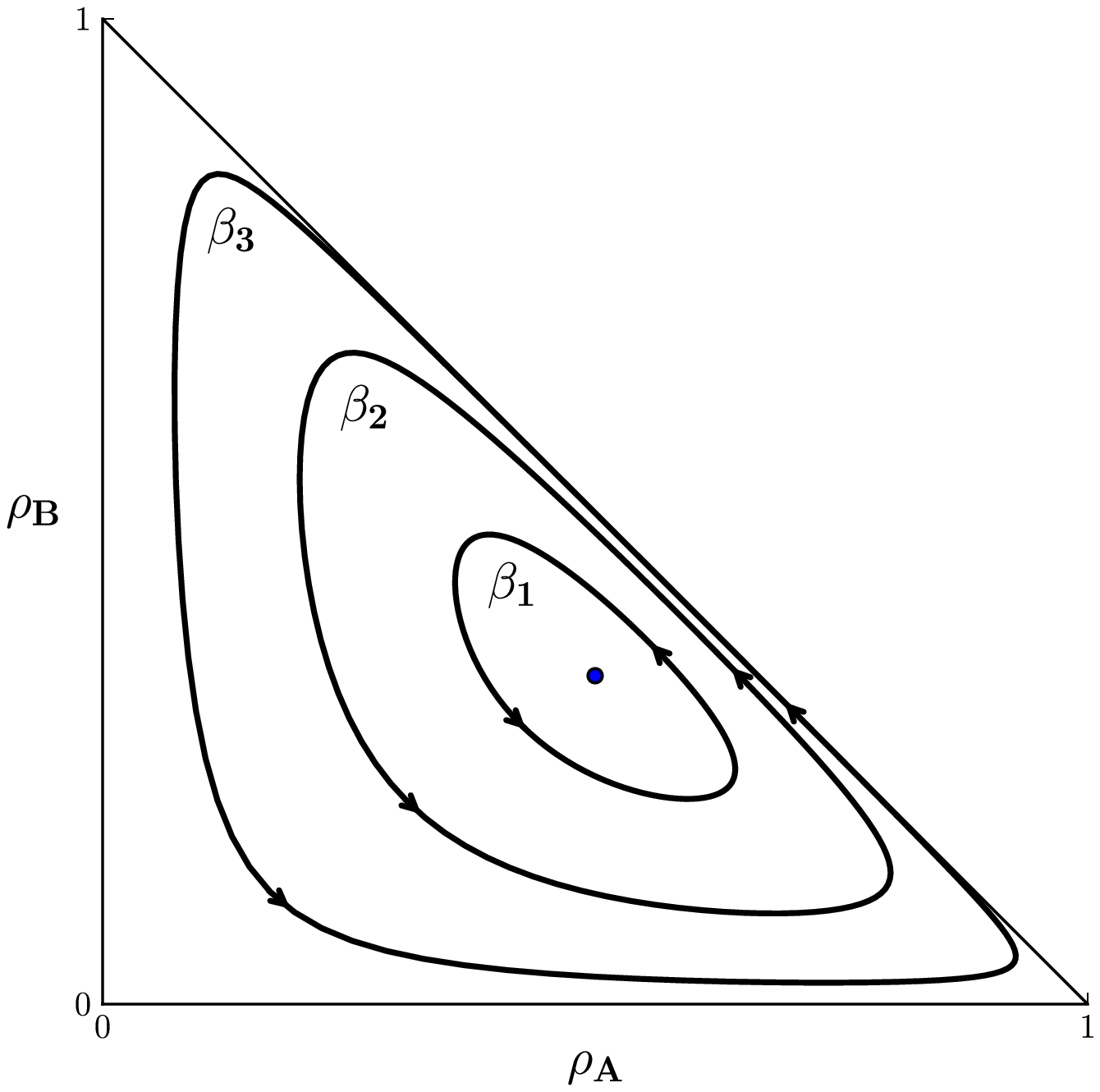}\hspace{-40pt}\includegraphics[scale=.5]{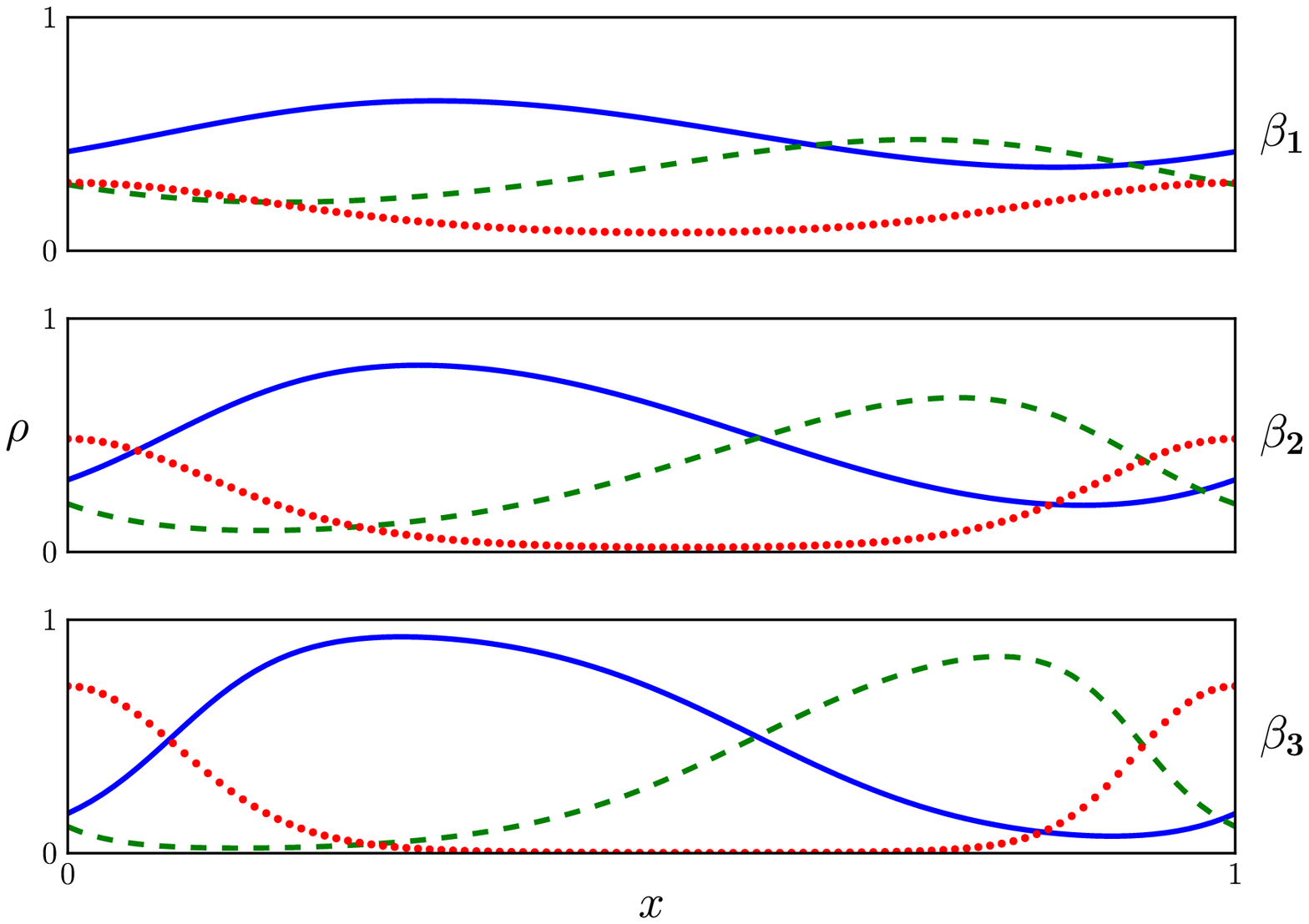}}
\caption{Plots of $\p_A$ (solid), $\p_B$ (dashed), and $\p_C$ (dotted), at right, and their corresponding trajectories in the $\p_A$-$\p_B$ plane, at left. See description under Region~I, Case~1.}
\label{fig:curves1}
\end{figure}

\begin{figure}
\centerline{\includegraphics[scale=.5]{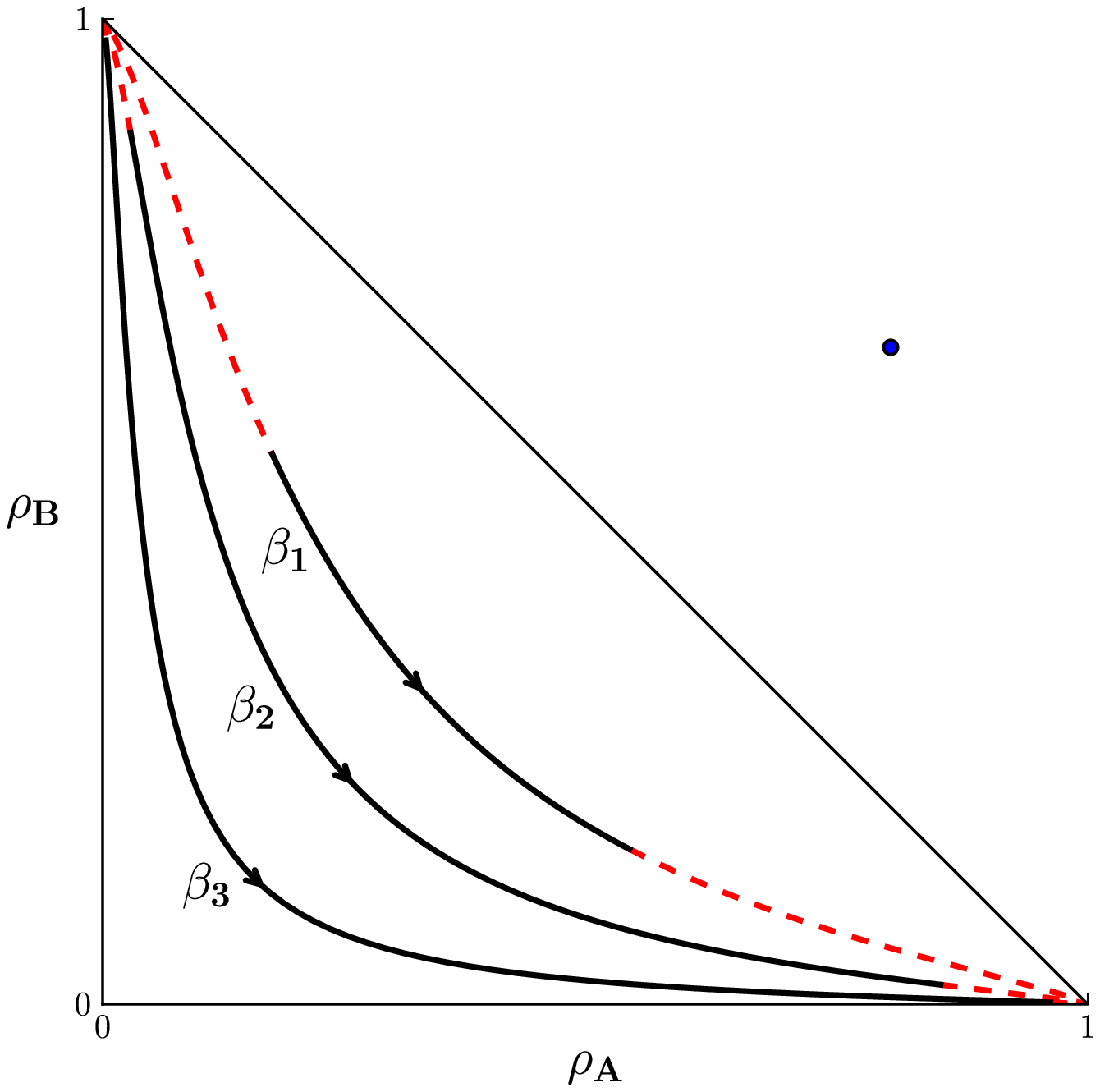}\hspace{-40pt}\includegraphics[scale=.5]{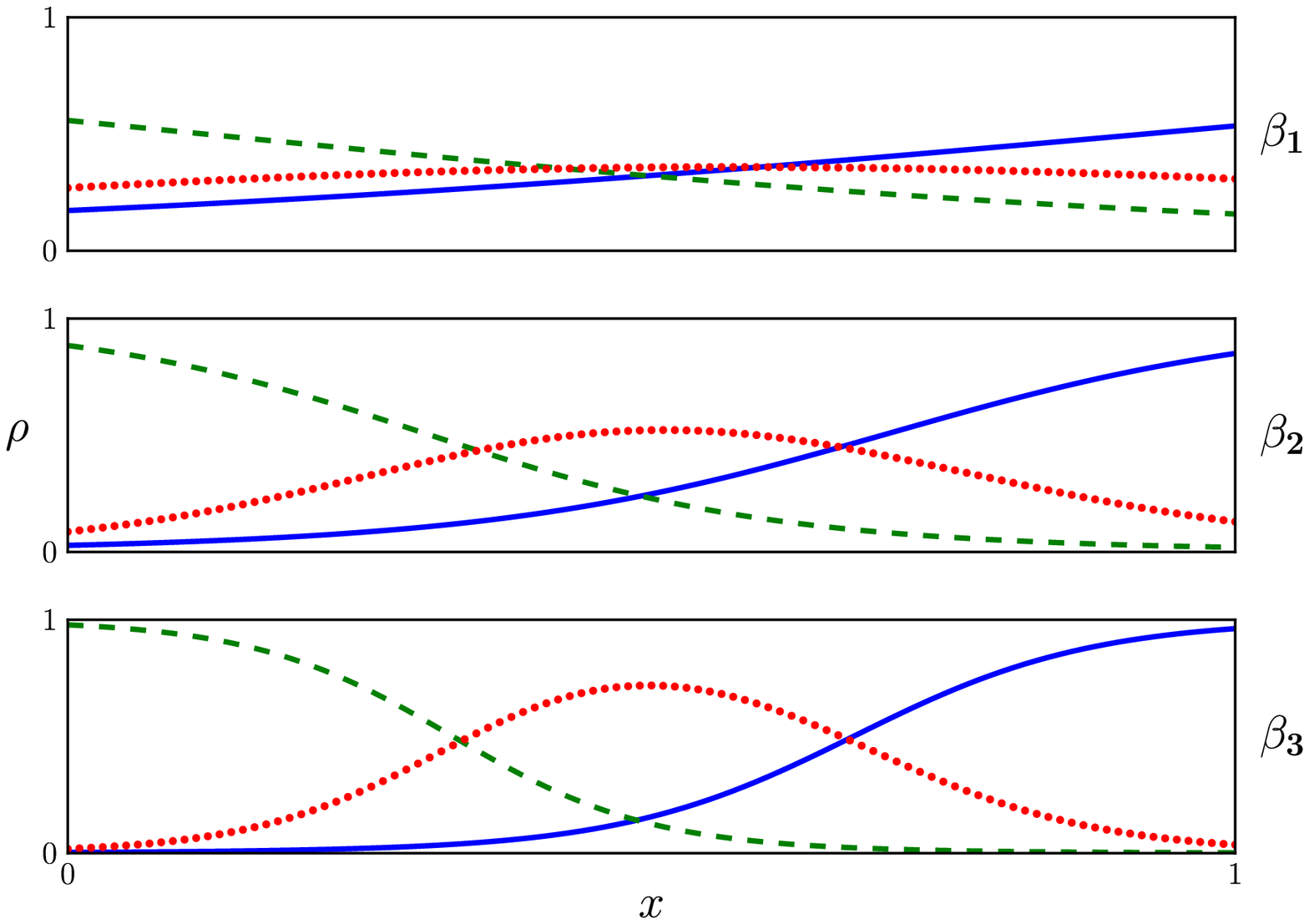}}
\caption{Plots of $\p_A$ (solid), $\p_B$ (dashed), and $\p_C$ (dotted), at right, and their corresponding trajectories in the $\p_A$-$\p_B$ plane, at left. See description under Regions~II and III.}
\label{fig:curves2}
\end{figure}

The trajectories of the densities $\p_\a(x)$ that are solutions of the ELE may be obtained by studying the level sets of the constant of the motion $K$ \eqref{K}, as described in Section~\ref{scalinglimit}. To do this let us define a line in the $\p_A$-$\p_B$ plane passing through the point $(v_A,\,v_B)$ by setting
\be \label{line}
\p_B=v_B+m\left(\p_A-v_A\right),
\ee
with $m$ an arbitrary constant. The change in $\log{K}$ as $\p_A$ is varied along the line \eqref{line} can be manipulated using \eqref{constraints} to yield
\bea
\frac{\hbox{d}\log{K}}{\hbox{d}\p_A} & = & \left(\frac{v_A}{\p_A}-\frac{v_C}{\p_C}\right)+m\left(\frac{v_B}{\p_B}-\frac{v_C}{\p_C}\right)\nn\\
                                    & = & \left[1+\frac{\p_A}{\p_C}\left(m^2+2m+1\right)+\frac{\p_A}{\p_B}m^2\right]\left(\frac{v_A}{\p_A}-1\right).
\label{Kline}
\eea
Thus $K$ is monotone increasing for $\p_A<v_A$ and decreasing for $\p_A>v_A$. A similar result holds for $\p_B$ and $v_B$. The shape of the level sets of $K$, and thus also the trajectories of the densities, depends upon which region the $v_\a$ lie in; compare Figures~\ref{fig:curves1} and \ref{fig:curves2}. We now give more details.

\medskip
\noindent{\bf\large Region I}\nopagebreak
\medskip

It follows from \eqref{Kline} that in region I $K$ achieves its maximum value at the center point $(v_A,\,v_B)$, where
\be \label{Kmax}
K\big|_{\p_\a=v_\a}=K_{\rm{max}}=\prod_\a{v_\a^{v_\a}},
\ee
and decreases monotonically as one moves along any straight line in the $(\p_A,\,\p_B)$ plane starting from the center. $K$ approaches its minimum value of zero on the boundaries of the triangle \eqref{triangle}, where one or more of the particle densities goes to zero. As $K$ is continuous in $\p_A$, $\p_B$ inside the triangle, this implies that the level sets of $K$ in this case consist of a single point at the center $(v_A,\,v_B)$ and closed curves surrounding the center point.

As the level curves of $K$ are closed, nonconstant solutions $\p_\a(x)$ of the ELE must be portions of periodic functions of $x$. When the period $\tau$ is greater (less) than $1$, the trajectory of the densities makes less (more) than one full rotation around the center point. We will refer to solutions with $n-1<1/\tau\leq n$, for $n$ an integer, as type $n$ solutions. For example, a solution that does not make one full rotation would be labeled as type $1$, while a solution making exactly three rotations around the center would be labeled as type $3$. This is consistent with the terminology used in \cconj{main}(a). The constant solution of the ELE is not assigned a type.

Trajectories of the particle densities satisfying the ELE for a given set of $v_\a$ depend upon the choice of the $r_\a$. For the $v_\a$ in region I there are two cases to consider: (1) $r_\a=v_\a$ for all $\a$, (2) $r_\a\neq v_\a$ for some $\a$.

\smallskip\noindent
{\bf Case 1: $r_\a=v_\a$.} In this case the model is rotation invariant, and one has $\p_\a(1)=\p_\a(0)$ for all $\a$. Here both constant and nonconstant solutions of the ELE are possible. The constant solution is given by $\p_\a(x)=v_\a$, $0\leq x\leq1$, as noted above. As $\p_\a(1)=\p_\a(0)$, see \eqref{BC}, nonconstant solutions must have an integer number of periods in the interval $x\in[0,1]$, corresponding to the number of times the trajectory orbits the center. Note that, as one moves along the interval in $x$, the maxima (and minima) of the particle densities proceed in cyclic order, that is, after species $\a$ reaches its maximum (minimum) density, the next species to achieve its maximum (minimum) density is $\a+1$. 

Example numerical solutions of the ELE in this case and their corresponding trajectories with $v_A=1/2$, $v_B=1/3$, $v_C=1/6$, and $r_\a=v_\a$ for all $\a$, are shown in Figure~\ref{fig:curves1}. In this plot the inverse temperatures are $\b_1=13$, $\b_2=15$, and $\b_3=20$, all larger than $\b_c$, which is $4\pi$ for this choice of the $v_\a$. The trajectories lie along lines of constant $K$, with $K_1\approx.349$, $K_2\approx.291$, and $K_3\approx.189$ for the solutions at $\b_1$, $\b_2$, and $\b_3$ respectively. Arrows indicate the flow along the trajectory as $x$ increases, and the point $(v_A,v_B)$ is marked by a dot.  For this case $K_{\rm max}\approx.364$.

\smallskip\noindent
{\bf Case 2: $r_\a\neq v_\a$.} Here there is no rotation invariance and the densities at opposite ends of the interval are not the same, $\p_\a(1)\neq\p_\a(0)$ for some $\a$, see \eqref{BC}. In this case only nonconstant solutions of the ELE are possible at finite temperatures, \ie for $\b>0$. These solutions will be portions of the periodic solutions described in Case~1. In contrast to the $r_\a=v_\a$ case, however, solutions of type $n$ do not exist for arbitrarily large values of $n$; there is some cutoff $n_{\rm{max}}\geq2$, which depends on the $r_\a$ and the $v_\a$, above which type $n$ solutions, $n\geq n_{\rm{max}}$, do not exist. This is because the average value of each density $\p_\a$ around one full orbit of the center is $v_\a$, so as $n$ becomes large the average density for the full profile $r_\a$ is steadily driven towards $v_\a$. See Appendix~\ref{excludedn} for more details.

\medskip
\noindent{\bf\large Regions II and III}
\medskip

When the $v_\a$ lie in regions II and III, the point $(v_A,\,v_B)$ lies outside the triangle \eqref{triangle}. Thus by \eqref{Kline} the level sets of $K$ inside the triangle cannot be closed curves, as when the $v_\a$ are in region I. Let us consider the value of $K$ along one of the boundaries of the triangle, where $\p_\a=0$ for some $\a$. If $v_\a$ is less than zero, $K$ will be infinite along this boundary. If $v_\a$ is zero $K$ will be finite, and if $v_\a$ is larger than zero $K$ will be zero. On the vertices where two boundaries meet, with $K$ equal to zero along one and infinite along the other, the value of $K$ at the vertex is not well defined, depending on the way in which the limit is taken. Thus when the $v_\a$ lie in regions II and III, lines of constant $K$ for $K$ finite and nonzero will be curves terminating on the vertices of the triangle where boundaries along which $K$ is infinite and zero meet. When the $v_\a$ lie on the boundaries between different regions, e.g. for the special values of the $v_\a$ considered in Section~\ref{specialcases}, the level sets of $K$ will be curves with ends terminating on either the edges or the vertices of the triangle. Solutions of the ELE are therefore nonconstant at finite temperatures and are not periodic.

Additionally, one may easily see by considering the ELE \eqref{ELE} that when one or two of the $v_\a$ are negative, the density of one particle species will monotonically increase in $x$, and another species will monotonically decrease. If for example $v_A<0$ while $v_B,v_C>0$, $\p_B$ will be monotonically increasing and $\p_C$ will be monotonically decreasing. Note that if the $v_\a$ lie in region II then as $x$ increases from $0$ to $1$ the maxima and minima of the particle densities proceed in cyclic order, as for region I systems, but when the $v_\a$ lie in region III the order is \textit{reversed}. That is, in region III a maximum (minimum) density of species $\a$ is followed by a maximum (minimum) of species $\a+2$.

Example numerical solutions and the corresponding trajectory of the densities with $v_A=4/5$, $v_B=2/3$, $v_C=-7/15$, $r_A=r_B=r_C=1/3$, are plotted in Figure~\ref{fig:curves2}. In this figure $\b_1=1$, $\b_2=3$, and $\b_3=5$. The trajectories lie along lines of constant $K$, with $K_1\approx.306$, $K_2\approx.167$, and $K_3\approx.066$ for the solutions at $\b_1$, $\b_2$, and $\b_3$ respectively. Portions of the level curves not traversed by the solutions are marked with dashed lines. Arrows indicate the flow along the trajectory, and the point $(v_A,v_B)$ is marked by a dot.

\section{The case $r_\a=v_\a$} \label{rvequal}

If $r_\a=v_\a$ for all $\a$ (so that we are necessarily in region~I) then as already noted the constant solution $\p_\a(x)=v_\a$, $0\le x\le1$, is always a solution of the ELE, and other solutions are given by traversing, exactly $n$ times for a type~$n$ solution, one of the simple closed curves which is a level set of $K$.  We show in Section~\ref{linearstability} that the constant solution cannot minimize the free energy when $\beta>\beta_c$, and in Section~\ref{typen} that no type~$n$ solution with $n\ge2$ can minimize the free energy.  Thus the existence of a unique minimizer as described in \cconj{main}(a) would be established if we could show that no non-constant solution can exist for $\beta<\beta_c$ and that for $\beta>\beta_c$ there is a unique type~1 solution, exactly what is proven in \cite{ACLMMS} for $v_\a=1/3$ for all $\a$.

Now under a change of scale $t=\beta x$ the ELE \eqref{ELE} become
\be \label{RSELE}
\frac{\hbox{d}\p_\a}{\hbox{d}t}
  =3\p_\a\left(v_{\a+1}\p_{\a+2}-v_{\a+2}\p_{\a+1}\right),
\ee
and type~1 solutions of \eqref{ELE} correspond to solutions of \eqref{RSELE} which have period $\beta$, the inverse of the temperature. The period $\tau(K)$ of the solution of \eqref{RSELE} is easily seen to be a continuous function of $K$ for $0<K\le K_{\rm max}$; a perturbative calculation as in \cite{ACLMMS} shows that  $\lim_{K\nearrow K_{\rm max}}\tau(K)=\beta_c$ as $K$ and we establish in Section~\ref{Kbeta} that $\lim_{K\searrow0}\tau(K)=\infty$.  The existence portion of \cconj{main}(a.ii) follows immediately from these observations.   If one could show that the period of the solution of \eqref{RSELE} is a monotonically decreasing function of $K$ for $0<K\le K_{\rm \max}$ then uniqueness would be established. However, one may establish a correspondence between the family of systems \eqref{RSELE} parameterized by the $v_\a$ in region~I and a large subset of the generalized Lotka-Volterra family of quadratic centers, specifically, those parameterized by $(b,\,c)$ with $b>0$ in the parameterization of \cite{V} (see Appendix~\ref{lvfamily} for details), and 
%However, \eqref{RSELE} is for arbitrary $v_\a$ a generic member of the generalized Lotka-Volterra family of quadratic centers \cite{V} (for details on the correspondence between \eqref{RSELE} and a standard parametrization of the Lotka-Volterra family, see Appendix~\ref{lvfamily}), and 
monotonicity of the period as a function of orbit size is for this family an open question \cite{V}.  In terms of the parameters used in \eqref{RSELE}, monotonicity was established in \cite{ACLMMS} when all $v_\alpha$ are $1/3$, and below we establish monotonicity when, for some $\a$, $v_\a=1/2$ and $v_{\a+1}=v_{\a+2}=1/4$.  Finally, monotonicity at sufficiently low temperature (or equivalently, small $K$) follows, for any $v_\a$, from the correspondence established in Appendix~\ref{lvfamily} and the results of \cite{V}.

\subsection{$v_{\a}=1/2$, $v_{\a+1}=v_{\a+2}=1/4$}

When the $v_\a$ lie in region I and two of them are equal, say $v_A=v_B=\gamma$ with $0<\gamma<1/2$, one may find explicitly an autonomous evolution equation for the third.  We begin with the expression
\be
K=\p_A(x)^{\gamma}\p_B(x)^{\gamma}\p_C(x)^{1-2\gamma},
\ee
which together with \eqref{constraints} may be used to obtain the densities $\p_A$ and $\p_B$ in terms of $\p_C$:
\bea\label{papb}
\p_A(x)=\frac{1}{2}\left(1-\p_C(x)\pm\sqrt{(1-\p_C(x))^2-4K^{1/\gamma}\p_C(x)^{-\lambda}}\right),\\
\p_B(x)=\frac{1}{2}\left(1-\p_C(x)\mp\sqrt{(1-\p_C(x))^2-4K^{1/\gamma}\p_C(x)^{-\lambda}}\right).
\eea
where $\lambda=(1-2\gamma)/\gamma$.  With \eqref{papb}  the reparameterized ELE \eqref{RSELE} for $\p_C$ is
\bea \label{CELE}
\p_C^{\prime}(t) & = & 3\gamma\p_C(t)\left(\p_B(t)-\p_A(t)\right)\nn\\
                & = & 3\gamma\p_C(t)\left(\pm\sqrt{(1-\p_C(t))^2-4K^{1/\gamma}\p_C(x)^{-\lambda}}\right).
\eea
Squaring \eqref{CELE} we have
\be \label{ELEenergyform}
\frac{9\gamma^2}{2}\p_C^{\prime}(t)^2+U_K(\p_C(t))=0,
\ee
with
\be \label{quarticpotential}
U_K(\p)=\frac{9}{32}\left(4K^{1/\gamma}\p^{2-\lambda}-\p^2\left(1-\p\right)^2\right).
\ee
This is the equation for a zero energy particle of mass $1$ confined in a  potential well. The zeros of the potential correspond to the turning points for the particle. Numerical calculations of the period, performed for many different values of $\gamma$, find that $\tau(K)$ is a monotonically decreasing function of $K$.

With $\gamma=1/4$ and thus $\lambda=2$ the potential becomes quartic, and an analytic calculation is possible.  Here the zeros of $U_K$ are
\be
\frac{1}{2}\pm\p_0,\qquad \p_0=\sqrt{\frac{1}{4}-2K^2}.
\ee
Note that in this case $K_{max}=1/2\sqrt{2}$, so $0<\p_0<1/2$. The period $\tau(K)$ of the type 1 solution $\p(t)$ may now be directly calculated, yielding
\bea
\tau(K) & = & \int_{1/2-\p_0}^{1/2+\p_0}{\! \text{d}\p \frac{2}{\sqrt{-2\,U_K(\p)}} } \nn\\\label{period}
     & = & \frac{16/3}{\sqrt{1/2-\p_0^2}-\p_0}F\left(\frac{\pi}{2},-\frac{4\p_0\sqrt{1/2-\p_0^2}}{\left(\sqrt{1/2-\p_0^2}-\p_0\right)^2}\right).
\eea
Here $F(\pi/2,\cdot)$ is the complete elliptic integral the first kind. The period \eqref{period} is a monotonically decreasing function of $K$, with (as expected) $\lim_{K\nearrow K_{max}}{\left(\tau(K)\right)}=8\sqrt{2}\pi/4=\b_c$ and $\lim_{K\searrow0}\tau(K)=\infty$. Thus for all $\b>\b_c$ there is a unique value of $K$ for which $\tau(K)=\beta$ and hence a unique type 1 solution of the ELE, and there is no type 1 solution for $\b<\b_c$.

\subsection{Linear stability of the constant solution for $\b<\b_c$} \label{linearstability}

We will now consider the linear stability of the constant solution $\p_\a(x)=v_\a$, $0\leq x\leq1$, via a computation similar to that of \cite{ACLMMS}. Let us consider two bounded continuous functions $\phi_A(x)$, $\phi_B(x)$, satisfying
\be
\int_0^1\!\hbox{d}x\;\phi_A(x)=\int_0^1\!\hbox{d}x\;\phi_B(x)=0,
\ee
and perturb the constant solution as
\be \label{perturbedrho}
(\p_A , \p_B , \p_C ) \rw (v_A + \epsilon\,\phi_A , v_B + \epsilon\,\phi_B , v_C - \epsilon\,(\phi_A + \phi_B))
\ee
for some fixed $\epsilon$ very small.  Under this perturbation all the terms in the free energy linear in $\epsilon$ cancel.  The order $\epsilon^2$ contribution to the entropy is
\be \label{svariation}
\frac{1}{2}\int_0^1\!\hbox{d}x\;\left[\frac{1}{v_A}\phi_A^2(x)+\frac{1}{v_B}\phi_B^2(x)+\frac{1}{v_C}(\phi_A(x)+\phi_B(x))^2\right].
\ee
The energy due to interaction with the external fields is linear in the densities, so the $\epsilon^2$ term for the energy is due entirely to the asymmetric mean field ABC interaction,
\be \label{evariation}
\begin{array}{l}
\displaystyle \b\int_0^1{ \!\hbox{d}x\int_0^1{ \!\hbox{d}y\;\Theta(y-x)\bigg[-\phi_A(x)(\phi_A(y)+\phi_B(y))+\phi_B(x)\phi_A(y)}}\\
\displaystyle \hskip140pt -(\phi_A(x)+\phi_B(x))\phi_B(y)\bigg]\\
\displaystyle \hskip40pt =3\b \int_0^1{ \!\hbox{d}x\int_0^1{ \hbox{d}y\;\Theta(y-x)\,\phi_B(x)\phi_A(y)}}.
\end{array}
\ee

Now expanding $\phi_A(x)$ and $\phi_B(x)$ in a Fourier series as
\bea
\phi_A(x)=\sum_{n=1}^\infty{(a_n\sin{(2\pi nx)}+b_n\cos{(2\pi nx)})},\\
\phi_B(x)=\sum_{n=1}^\infty{(c_n\sin{(2\pi nx)}+d_n\cos{(2\pi nx)})},
\eea
we see that the second variation of the free energy functional around the constant solution is given by
\be
\begin{array}{l}
\displaystyle \sum_{n=1}^\infty{\bigg[\frac{v_A+v_C}{v_A v_C}\,(a_n^2+b_n^2)+\frac{v_B+v_C}{v_B v_C}\,(c_n^2+d_n^2)}\\
\displaystyle \hskip80pt +\frac{2}{v_C}(a_n c_n + b_n d_n) -\frac{3\b}{\pi n}(b_n c_n - a_n d_n)\bigg].
\end{array}
\ee
If we write $u_n=(a_n,\; b_n, \;c_n, \;d_n)$, the second variation can be expressed in matrix form as $\sum_{n=1}^\infty{u_n\,M\,u_n^T}/\left(v_Av_Bv_C\right)$, where
\be
M = \left[\begin{matrix} v_B(1-v_B)           & 0                     & v_A v_B               & 3\b/(2\pi n)         \\
                        0                    & v_B(1-v_B)            & -3\b/(2\pi n)         & v_A v_B              \\
                        v_A v_B              & -3\b/(2\pi n)         & v_A(1-v_A)            & 0                    \\
                        3\b/(2\pi n)         & v_A v_B               & 0                     & v_A(1-v_A)           \\
         \end{matrix}\right].
\ee
The eigenvalues of $M$ are
\be
\begin{array}{l}
\displaystyle \lambda_\pm = \frac{1}{2}\Bigg[v_A(1-v_A)+v_B(1-v_B)\\
\displaystyle \hskip40pt \pm\sqrt{\left(v_A(1-v_A)+v_B(1-v_B)\right)^2-4\left(v_Av_Bv_C-\left(\frac{3\b}{2\pi n}\right)^2\right)}\Bigg],
\end{array}
\ee
each with degeneracy $2$. Thus for $\b\leq\b_c=2\pi/(3\sqrt{v_A v_B v_C})$ the matrix $M$ is positive definite, so the second variation of the free energy around the constant solution is also positive and the constant solution is a local minimum. At lower temperatures, $\b>\b_c$, the smallest eigenvalue $\lambda_-$ becomes negative, so in this regime the constant solution can no longer be the minimizer of the free energy.

\subsection{Solutions of type $n\geq2$} \label{typen}

In this section we will show that solutions of the ELE for $r_\a=v_\a$ with two or more full periods in $x\in[0,1]$ can never minimize the free energy.

Consider a set of profiles $\{\p\}$, not necessarily solutions of the ELE but satisfying the constraints \eqref{constraints}.  Then for any integer $n\geq2$ we define the set of profiles $\{\hat{\p}_n\}$ by
\be
\hat{\p}_{n,\a}(x)=\p_\a(nx-j)\;\;\;\;\hbox{for }\frac{j}{n}\leq x\leq \frac{j+1}{n},\;\;\;\;j=0,\ldots,n-1.
\ee
That is, we obtain the $\hat{\p}_{n,\a}$ by shrinking the $\p_\a$ horizontally by a factor of $n$ and repeating these reduced profiles $n$ times in the interval $[0,1]$.  We claim then that
\bea
\cS(\{\hat{\p}_{n}\})&=&\cS(\{\p\}),\\\label{Econst}
\cE(\{\hat{\p}_{n}\})&=&\frac{1}{n}\cE(\{\p\})+\left(1-\frac{1}{n}\right)\cE(\{r\}),
\eea
where $\cE(\{r\})=9v_Av_Bv_C/2$ is the energy of the constant solution $\p_\a(x)=v_\a$.

The proof proceeds by direct calculation. First let us consider the entropy of the new profiles,
\bea
\cS(\{\hat{\p}_{n}\})&=&-\sum_{\a}{ \sum_{j=0}^{n-1}{ \int_{j/n}^{(j+1)/n}{ \hbox{d}x\, \hat{\p}_{n,\a}(x)\log{\hat{\p}_{n,\a}(x)}}}}\nn\\
&=&-\sum_{\a}{ \sum_{j=0}^{n-1}{ \frac{1}{n}\int_0^1{\hbox{d}\tilde{x}\,\p_\a(\tilde{x})\log{\p_\a(\tilde{x})}}}}=\cS(\{\p\}).
\eea
Here we have used a change of variables with $x=(\tilde{x}+j)/n$.  The calculation of the energy proceeds similarly.  Using the same change of variables we have
\bea \label{Erhohat}
\cE(\{\hat{\p}_{n}\})&=&3\sum_{\a}{\sum_{j,k=0}^{n-1}{\frac{1}{n^2}\int_0^1{\!\hbox{d}\tilde{x}}\int_0^1{\!\hbox{d}\tilde{y}\;\Theta\!\left(\frac{\tilde{y}+k-\tilde{x}-j}{n}\right)v_{\a+1}\p_\a(\tilde{x})\p_{\a+2}(\tilde{y})}}}\nn\\
&=&3\sum_{\a}{\Bigg\{\sum_{j=0}^{n-1}{\frac{1}{n^2}\int_0^1{\!\hbox{d}\tilde{x}\int_0^1{\!\hbox{d}\tilde{y}\;\Theta\!\left(\tilde{y}-\tilde{x}\right)v_{\a+1}\p_\a(\tilde{x})\p_{\a+2}(\tilde{y})}}}}\nn\\
& &\;\;\;+\sum_{0\leq j<k\leq n-1}{\frac{1}{n^2}\int_0^1{\!\hbox{d}\tilde{x}\int_0^1{\!\hbox{d}\tilde{y}\;v_{\a+1}\p_\a(\tilde{x})\p_{\a+2}(\tilde{y})}}}\Bigg\}\nn\\
&=&\frac{1}{n}\cE(\{\p\})+\left(1-\frac{1}{n}\right)\cE(\{r\}),
\eea
as there are $n(n-1)/2$ pairs of indices with $0\leq j<k\leq n-1$.

Now there are two cases to consider.  If $\cE(\{\p\})<\cE(\{r\})$, then by \eqref{Erhohat} $\cE(\{\hat{\p}_{n}\})>\cE(\{\p\})$, thus the original profiles $\p_\a$ have a lower free energy and thus the $\hat{\p}_{n,\a}$ cannot be minimizers.  If instead $\cE(\{\p\})\geq\cE(\{r\})$, then $\cE(\{\hat{\p}_{n}\})\geq\cE(\{r\})$ as well, and as long as $\p_\a$ is not the constant solution $\cS(\{r\})>\cS(\{\hat{\p}_{n}\})$. Then $\cF(\{r\})<\cF(\{\hat{\p}_{n}\})$, so again the $\hat{\p}_{n,\a}$ are not minimizers.  A type $n$ solution of the ELE at an inverse temperature $\b$ is of the form $\hat{\p}_{n,\a}$, where $\p_\a$ is a type $1$ solution at an inverse temperature $\b/n$.  Thus no type $n$ solution for $n\geq2$ can minimize the free energy.

\subsection{$K$-$\b$ relation}\label{Kbeta}

It follows from the result of  Section~\ref{typen} that when $r_\a=v_\a$ a
minimizer of the free energy must be either the constant solution of the
ELE or a type 1 solution. Let  $K(\beta)$ denote  the value of $K$ for the
minimizer at temperature $\beta$; if several minimizers exist then we
choose one of them arbitrarily to define $K(\beta)$.  Of course, if
$K(\beta)<K_{\rm max}$ then $\tau(K(\beta))=\beta$.  If $\tau(K)$ is
monotonic then $K(\b)$ will be the inverse of the function $\tau(K)$ and must
be continuous, since $\tau(K)$ is, but we cannot show this and thus cannot
rule out the possibility that  $K(\beta)$ may be discontinuous.  However,
we do show here  that $K(\beta)$ must be monotonic decreasing.

We begin with  the integral form \eqref{ELEint} of the ELE:
\be
\log{\p_\a(x)}=\log{\p_\a(0)}+3\b\int_0^x{\! \text{d}y \left(v_{\a+1}\p_{\a+2}(y)-v_{\a+2}\p_{\a+1}(y)\right)}.
\ee
Substituting this into the entropy \eqref{macroentropy} and using $\int_0^1\p_\a(x)\,dx=r_\a$ gives
\begin{align}
\sum_\a{r_\a\log{\p_\a(0)}} &= -3\b\sum_\a{\int_0^1{\!\text{d}x \int_0^x{\!\text{d}y\; \p_\a(x)\left(v_{\a+1}\p_{\a+2}(y)-v_{\a+2}\p_{\a+1}(y)\right)}}}\nn\\
     & \hskip80pt -\cS(\{\p\})\nn \\
                      &= 6\b\sum_\a{\int_0^1{\!\text{d}x \int_0^1{\!\text{d}y\; \Theta(y-x)v_{\a+1}\p_\a(x)\p_{\a+2}(y)}}}\nn\\
                           & \hskip80pt -\cS(\{\p\})-3\b\sum_\a{v_\a r_{\a+1} r_{\a+2}}\nn\\\label{logK}
                           &= 2\b\cE(\{\p\})-\cS(\{\p\})-2\b\cE(\{r\}),
\end{align}
where $\cE(\{r\})$ (see \eqref{energy}) is the energy of the constant profile $\p_\a(x)=r_\a$ (which is  not a solution of the ELE unless $r_\a=v_\a$). \eqref{logK} is a general relation which holds for all profiles $\{\p\}$ satisfying the ELE, \ie for all stationary points of $\cF(\{\p\})$, whether or not $r_\a=v_\a$.

Now suppose that  $r_\a=v_\a$ for all $\a$. The left hand side of \eqref{logK} is then just $\log{K}=\sum_\a{v_\a\log{\p_\a(x)}}$, the same for all $x\in[0,1]$. Then if $\{\rho\}=\{\rho^{(\b)}\}$ is the minimizing solution of the ELE corresponding to $K(\beta)$, \eqref{logK} becomes
\be\label{logKeq}
\log K(\beta) = 2\b\cE(\{\p^{(\b)}\})-\cS(\{\p^{(\b)}\})-2\b\cE(\{r\}),
\ee
where now $\cE(\{r\})=\cE(\{\p^{(0)}\})=9v_A v_B v_C/2$. For $\beta_2>\beta_1$ we subtract the corresponding equations \eqref{logKeq} and rearrange terms to obtain
\begin{align}\nn
\log\frac{K(\b_2)}{K(\b_1)}
  &=2\bigl(\cF_{\b_2}(\{\p^{(\b_2)}\})-\cF_{\b_2}(\{\p^{(\b_1)}\})\bigr)
 +\bigl(\cS(\{\p^{(\b_2)}\})-\cS(\{\p^{(\b_1)}\})\bigr)\\
 &\hskip50pt\label{final}
  +2(\b_2-\b_1)\bigl(\cE(\{\p^{(\b_1)}\})-\cE(\{\p^{(0)}\})\bigr).
\end{align}
where we have indicated the explicit $\beta$ dependence in \eqref{energy} by writing $\cF_\b$.  But it follows from simple general thermodynamic arguments that both $\cS(\{\p^{(\b)}\})$ and $\cE(\{\p^{(\b)}\})$ are monotonic decreasing functions of $\b$, and since $\p^{(\b_2)}$ minimizes $\cF_{\b_2}$ all three terms on the right side of \eqref{final} are nonpositive.  This establishes the monotonicity of $K(\b)$. We note in addition that, if at a particular value of the temperature $\b$ there exist several minimizers of the free energy, $\{\p_i^{(\b)}\}$, $i=1,\ldots,n$, then $\log{(K_i/K_j)}=\cS(\{\p_i^{(\b)}\})-\cS(\{\p_j^{(\b)}\})$.

One may also see that when $\b$ is large, $K(\b)$ must be small. For from
\eqref{logKeq} we have
\be \label{betalogk}
\frac{\log{K(\b)}}{\b}
 =2\left(\cE(\{\p^{(\b)}\})-\cE(\{r\})\right)+\b^{-1}\cS(\{\p^{(\b)}\});
\ee
since the energy and entropy are bounded functions and $\cE(\{\p^{(\b)}\})$ is decreasing in $\beta$ the right hand side  of \eqref{betalogk} approaches a finite value as $\beta\nearrow\infty$. In fact, one can show that $\lim_{\beta\nearrow\infty}\cE(\{\p^{(\b)}\})$ is the ground state energy per particle $3v_Av_Bv_C$ (see Appendix~\ref{microsystem}), so that asymptotically $\log K(\b)\sim-(3/2)v_Av_Bv_C\b$.

%In fact, one can show that $\lim_{\beta\nearrow\infty}\cE(\{\p^{(\b)}\})$ is the ground state energy per particle (see Appendix~\ref{microsystem}). The difference in energies in \eqref{betalogk} then goes to $\cE(\{\p^{(\infty)}\})-\cE(\{r\})=3v_Av_Bv_C-9v_A v_B v_C/2=-(3/2)v_Av_Bv_C$, so that asymptotically $\log K(\b)\sim-(3/2)v_Av_Bv_C\b$.

\section{Special cases when $r_\a\neq v_\a$, with the $v_\a$ outside of region~I} \label{specialcases}

For certain values of the $r_\a$ and $v_\a$ we are able to prove \cconj{main}(b). There are two such special cases. In the first case, one of the $v_\a$ is zero and $v_{\a\pm1}$ have opposite signs (so that one of $v_{\a\pm1}$ must be greater than one). In this case, the $v_\a$ lie in fact on the \textit{boundary} between regions II and III. In the second case one of the $v_\a$ is one. We will present the proof of \cconj{main}(b) for these cases in the sections below.

\subsection{$v_\a=0$, $v_{\a\pm1}>1$}

For definiteness let us take $v_C=0$, $v_B>1$. On the orbit with conserved quantity $K$ we have $\rho_A(x)^{v_A}\rho_B(x)^{1-v_A}=K$ and so
\be\label{rhoAB}
\rho_A(x)=K^{1/v_A}\rho_B(x)^\gamma,\qquad \rho_B(x)=K^{-1/(v_A-1)}\rho_A(x)^{1/\gamma},
\ee
where $\gamma=(v_A-1)/v_A$ satisfies $\gamma>1$. Thus the profiles satisfy
\begin{align}\label{rhoA}
\rho_A'&=3\beta\rho_A(v_B\rho_C)
=3\beta (1-v_A) \rho_A(1-\rho_A-K^{1/(1-v_A)}\rho_A^{1/\gamma})
\equiv f_K(\rho_A),\\\label{rhoB}
\rho_B'&=3\beta\rho_B(-v_A\rho_C)
=-3\beta v_A\rho_B(1-\rho_B-K^{1/v_A}\rho_B^{\gamma})
\equiv g_K(\rho_B).
\end{align}
Note that if $K^*>K$ then
\be\label{signs}
f_{K}(\rho)>f_{K^*}(\rho)>0 \qquad\hbox{and}\qquad g_{K^*}(\rho)>g_K(\rho)>0.
\ee

Now consider two profiles $\rho_\alpha(x)$ and $\rho_\alpha^*(x)$ satisfying the ELE which have different starting values: $(\rho_A(0),\rho_B(0))\ne(\rho_A^*(0),\rho_B^*(0))$; we claim that the corresponding averages are not equal: $(r_A,r_B)\ne(r_A^*,r_B^*)$.  We denote the corresponding conserved quantities by $K$ and $K^*$ and consider several cases.

\smallskip\noindent
{\bf Case 1: $K=K^*$.} Without loss of generality we take $\rho_A^*(0)>\rho_A(0)$ and hence by \eqref{rhoAB} also $\rho_B^*(0)>\rho_B(0)$. Since $\rho_\alpha(x)$ and $\rho_\alpha^*(x)$, $\alpha=A,B$, satisfy the same differential equation which by \eqref{signs} has a strictly negative right hand side we have $\rho_\alpha(x)>\rho_\alpha^*(x)$ for all $x$ and so 
\be\label{case1}
r_A^*>r_A \qquad\hbox{and}\qquad r_B^*>r_B.
\ee

\smallskip\noindent
{\bf Case 2: $\rho_A^*(0)=\rho_A(0)$.} Now without loss of generality we may take $K^*>K$ and so by \eqref{rhoAB} we have $\rho_B^*(0)>\rho_B(0)$.  Thus by \eqref{rhoB} and \eqref{signs} we have that $\rho_B^*(x)>\rho_B(x)$, and by \eqref{rhoA} and \eqref{signs} that $\rho_A(x)>\rho_A^*(x)$ for $0< x\le1$, and so
\be\label{case2}
r_A^*<r_A \qquad\hbox{and}\qquad r_B^*>r_B.
\ee

\smallskip\noindent
{\bf The general case:} Since we have dealt with the possibility $K^*=K$ in Case~1 we may take $K^*>K$.  The case $\rho_A^*(0)=\rho_A(0)$ has
been considered in Case~2.  If $\rho_A^*(0)>\rho_A(0)$ we introduce a profile $\rho_\alpha^{**}$ with $\rho_A^{**}(0)=\rho_A^*(0)$ and $K^{**}=K$; then by Cases~1 and 2 we have
\be
r_B^*>r_B^{**}>r_B.
\ee
If instead $\rho_A^*(0)<\rho_A(0)$ we argue similarly, introducing a profile $\p_\a^{**}$ with $\rho_A^{**}(0)=\rho_A(0)$ and $K^{**}=K^*$, obtaining
\be
r_A>r_A^{**}>r_A^*.
\ee
Thus the solution of the ELE must be unique.

The $v_C=0$, $v_B<0$, $v_A=1-v_B>1$ case may be argued very similarly, but making use of comparisons between the values of the densities at the end of the interval $\p_\a(1)$ rather than the initial values $\p_\a(0)$, $\alpha=A,B$. Otherwise the argument proceeds identically to the case above, so we will omit the full derivation here.

\subsection{$v_\a=1$}

When one of the $v_\a=1$ it is possible to solve the ELE exactly. Let us assume $v_A=1$ and $v_B$, $v_C$ are not zero, a trivial case, so the $v_\a$ lie in region II. The ELE for $\p_A$ and $\p_B$ become
\bea
\frac{\hbox{d}\p_A}{\hbox{d}x} & = & 3\b v_B\,\p_A(\p_B+\p_C)\nn\\
                              & = & 3\b v_B\,\p_A(1-\p_A),\\
\frac{\hbox{d}\p_B}{\hbox{d}x} & = & 3\b \p_B(v_C\,\p_A-\p_C)\nn\\
                              & = & -3\b\,\p_B(1-\p_B)+3\b(1-v_B)\p_A\,\p_B. 
\eea
The equation for $\p_A$ may easily be solved to obtain
\be \label{rhoasolution}
\p_A(x)=\frac{1}{1+c_A\,e^{-3\b v_B\,x}},
\ee
where $c_A=1/\p_A(0)-1$. Using \eqref{rhoasolution} one may then solve for the density $\p_B$,
\be \label{rhobsolution}
\p_B(x)=c_A\left(c_A+e^{3\b v_B\,x}\right)^{\frac{1-v_B}{v_B}}\frac{1}{\left(1+c_A\right)^{1/v_B}+c_A\,c_B\,e^{3\b v_B\,x}},
\ee
with the constant
\bea
c_B & = & \left(1+c_A\right)^{1/v_B}\left(\frac{1}{(1+c_A)\p_B(0)}-\frac{1}{c_A}\right)\nn
%    & = & \frac{1-\p_A(0)-\p_B(0)}{c_A\,\p_B(0)\,\p_A(0)^{1/v_B}}.
\eea

To show that the solution is unique, we must demonstrate that there is only one choice of the initial values $\p_\a(0)$ which will yield a particular set of average densities $r_\a$ at a given $\b$. This may be seen directly from the density profiles \eqref{rhoasolution}, \eqref{rhobsolution}, which depend simply on $c_A$ and $c_B$.  It is easy to see that
\be
\frac{\hbox{d}\p_A(x)}{\hbox{d}c_A}<0,\qquad \frac{\hbox{d}\p_B(x)}{\hbox{d}c_B}<0,
\ee
and as $\hbox{d}c_A/\hbox{d}\p_A(0)<0$, $\hbox{d}c_B/\hbox{d}\p_B(0)<0$, we have
\be
\frac{\hbox{d}r_A}{\hbox{d}\p_A(0)}>0,\qquad \frac{\hbox{d}r_B}{\hbox{d}\p_B(0)}>0.
\ee
Thus there is a unique solution of the ELE, given by \eqref{rhoasolution}, \eqref{rhobsolution}, for all $\b$. The average density of the profiles is given by
\bea
\label{rAsolution} r_A & = & \frac{1}{3\b v_B}\log{\frac{e^{c_A}+e^{3\b v_B}}{1+e^{c_A}}},\\
\label{rBsolution} r_B & = & 1-\frac{1}{3\b}\log{\frac{\left(e^{c_A}+e^{3\b v_B}\right)^{1/v_B}+e^{3\b+c_A}\,c_B}{\left(1+e^{c_A}\right)^{1/v_B}+e^{c_A}\,c_B}},\\
r_C & = & 1-r_A-r_B.
\eea
In principle one may invert equations \eqref{rAsolution}, \eqref{rBsolution} to write the solutions in terms of the average densities $r_A$ and $r_B$ rather than $\p_A(0)$ and $\p_B(0)$.

\bigskip
\noindent {\normalsize {\bf Acknowledgments} We thank Lorenzo Bertini, Thierry Bodineau, Bernard Derrida, David Mukamel, and Oliver Penrose for useful discussions. The work of J.B. and J.L.L. was supported in part by NSF grant DMR-08-02120 and AFOSR grant FA-95550-10-1-0131.}

\appendix

\section{Properties of the microscopic system} \label{microsystem}

Many of the properties of the canonical measure $\mu_\b$ for the standard ABC model, described in \cite[Sections~1 and 2]{ACLMMS}, hold  with relatively minor changes for the generalized model studied here. Below we discuss a few of the necessary modifications.

As in the standard ABC model, there is a certain nearest neighbor exchange dynamics which satisfies detailed balance with respect to the canonical Gibbs measure $\mu_\b$. In this dynamics a particle of type $\a$ at site $i$ and a particle of type $\gamma$ at site $i+1$ exchange places $\a\,\gamma\,\rw\,\gamma\,\a$ with rate $q_{\a\gamma}$,
\be \label{rates}
q_{\a\gamma}=\left\{\begin{gathered}
                        e^{-3\b\, v_{\a+2}/N}\;\;\;\;\;\;\hbox{if }\gamma=\a+1\hfill \\
                        1\;\;\;\;\;\;\;\;\;\;\;\;\;\;\;\;\;\;\;\;\;\;\hbox{if }\gamma=\a-1\hfill \\
                       \end{gathered} \;,
\right.
\ee
where the $v_\a$ are as in \eqref{valpha}. When the $\xi_\a$ are not all zero (\ie the $v_\a$ are not all equal), the system no longer has cyclic symmetry in the particle types.

In analogy with what happens in the standard ABC model when $N_\a=N/3$ for all $\a$, we have here a rotation invariant energy when $v_\a\,N=N_\a$ for all $\a$ (this is clearly only possible when the $v_\a$ lie in region I). That is, if one imagines connecting site $N$ to site $1$ and then rotating the configuration $\underline{\eta}$, the rotation leaves $E$ unchanged. A simple way to check the rotation invariance is to consider moving a particle of type $\a$ from the end of the interval at site $N$ to site $1$, and translating all the other particles from sites $i$ to $i+1$.  The change in energy after this rotation is then $3\left(v_{\a+1}\,N_{\a+2}-v_{\a+2}\,N_{\a+1}\right)/N$, which vanishes for $v_\a\,N=N_\a$. Note that in this rotation invariant case the rates \eqref{rates} for a particle of type $\a$ at a site $i$ and a particle of type $\gamma$ at $i+1$ to exchange become
\be
q_{\a\gamma}=\left\{\begin{gathered}
                        e^{-3\b\, N_{\a+2}/N^2}\;\;\;\;\;\;\hbox{if }\gamma=\a+1\hfill \\
                        1\;\;\;\;\;\;\;\;\;\;\;\;\;\;\;\;\;\;\;\;\;\;\;\;\hbox{if }\gamma=\a-1\hfill \\
                       \end{gathered} \;,
\right.
\ee
which satisfies the general condition on exchange rates derived in \cite{EKKM2} for which detailed balance holds on the ring.

We remark that the energy in this rotation invariant case may be constructed in a different way, beginning from the standard ABC model energy with $\xi_\a=0$.  If one wishes to write down an ABC-like energy that is explicitly rotation invariant, one way to achieve this is to take the standard ABC energy and average over starting the sum at each point of the lattice, \ie
\be \label{rimicroenergy}
\tilde{E}(\underline{\eta})=\frac{1}{N}\sum_{k=1}^N{ \left( \frac{1}{N}\sum_{\a}{ \sum_{i=1+k}^{N+k}{ \sum_{j=1+k}^{N+k}{ \Theta(j-i)\eta_\a(i)\eta_{\a+2}(j) } } } \right) },
\ee
where one imagines the interval with periodic boundary conditions, such that site $N+m$ refers to site $m$.  This gives
\be
\tilde{E}(\underline{\eta})=\frac{1}{N}\sum_{\a}{ \sum_{i=1}^{N}{ \sum_{j=1}^{N}{ \left(\Theta(j-i)+\frac{i-j}{N}\right)\eta_\a(i)\eta_{\a+2}(j) } } },
\ee
where the interaction term depends upon the distance between sites $i$ and $j$ as well as their order on the line.  One may easily rearrange this expression to find
\be
\tilde{E}(\underline{\eta}) = \frac{1}{N}\sum_{\a}{ \sum_{i=1}^N{ \sum_{j=1}^N{  \Theta(j-i)\,3\,\frac{N_{\a+1}}{N}\,\eta_\a(i)\,\eta_{\a+2}(j) } } }, 
\ee
which is identical to \eqref{microenergy2} up to a constant when $v_\a\,N=N_\a$.

Now let us consider the ground states of the model. As in the standard ABC model, in the $\b\rw\infty$ limit the particle species become phase separated, with the ground states consisting of macroscopic domains of pure $A$, $B$, and $C$ particles.  The arrangement of these domains may depend upon the $v_\a$ as well as the number of each particle species $N_\a$.  For values of $v_\a$ in regions II and III, where one or two of the $v_\a$ is negative, the ground state is completely determined by requiring that all nearest neighbor configurations be stable, i.e., that the energy may not be lowered by making a nearest neighbor exchange.  Consider the possible orderings of nearest neighbor pairs of different particle species: $AB$, $AC$, $BA$, $BC$, $CA$, $CB$.  When the $v_\a$ do not all have the same sign, the energetically preferred alignment of two nearest neighbor particles of different species will be cyclic for some pairs and anti-cyclic for others. Thus it is not possible to have an arrangement of four or more domains where all nearest neighbor pairs are preferably aligned.  Generally, when the $v_\a$ lie in region II (III), if $v_\gamma<0\;(>0)$, the ground state is given by three domains arranged in cyclic (anti-cyclic) order, \ie in a cyclic (anti-cyclic) permutation of $\mathbf{ABC}$, with the domain of type $\gamma$ particles in the middle.  For example, if $v_A<0$ while $v_B,\,v_C>0$, the stable nearest neighbor configurations are $AB$, $CB$, and $CA$, thus the ground state must consist of three domains of particles cyclically arranged as $\mathbf{CAB}$.

When the $v_\a$ lie in region I, the ground states of the ABC model with external fields are closely related to those of the standard ABC model.  To show the correspondence, let us consider a rescaling of the energy \eqref{microenergy2}, $\tilde{E}(\eta)=E(\eta)/(v_A\,v_B\,v_C)$.  The rescaled energy of a sequence of domains, where the $i$th domain consists of $k_{\a,i}$ particles of type $\a$ (and no particles of any other type), is up to a constant factor formally identical to the standard ABC energy of the same sequence of domains, but with $k_{\a,i}/v_\a$ particles in each domain.  Of course, $k_{\a,i}/v_\a$ may not be an integer, which must be taken into account when considering the degeneracy of ground states.  One may however apply the same analysis as in \cite{ACLMMS} to determine the lowest energy configuration of domains. This ground state energy per site is given, up to a constant term (see \eqref{microenergy2}), by $3\min{\{v_A\,r_B\,r_C,\,v_B\,r_C\,r_A,\,v_C\,r_A\,r_B\}}$. One then finds that the ground state of the generalized ABC model, with $v_\a$ in region I and $N_\a$ particles of type $\a$, is arranged in the same order and has the same symmetries as the ground state of the standard ABC model with $N_\a/v_\a$ particles of type $\a$. There are three different cases.
\begin{enumerate}
\item If one of the $N_\a/v_\a$ is larger than the other two then there exists a unique ground state with three domains arranged in cyclic order, with the particles for which $N_\a/v_\a$ is largest in the middle. For example if $N_A/v_A>N_{B,C}/v_{B,C}$, the ground state arrangement of domains is $\mathbf{CAB}$, with the particles of type $A$ in the middle domain.
\item If two of the $N_\a/v_\a$ are equal and larger than the term for the third species, the ground state is degenerate, consisting of three or four domains cyclically arranged with the particles for which $N_\a/v_\a$ is smallest placed on the boundaries of the interval. If for instance $N_{B}/v_{B}=N_{C}/v_{C}>N_A/v_A$, the ground state is $N_A+1$ degenerate, with the domains arranged as $\mathbf{ABC}$, $\mathbf{ABCA}$, or $\mathbf{BCA}$; the type $A$ particles may appear on either the left or the right sides of the interval.
\item If all of the $N_\a/v_\a$ are equal, the ground state is $N$ degenerate, consisting of arbitrary rotations of the domains arranged in $\mathbf{ABC}$ cyclic order. In this case, as discussed above, the energy of any configuration $\underline{\eta}$ is invariant under rotation.
\end{enumerate}

\section{The Lotka-Volterra family of centers and ABC-like systems of ODEs} \label{lvfamily}

In this section we will demonstrate that the ABC model with external fields is a member of the Lotka-Volterra family of ODE systems when $0<v_\a<1$ for all $\a$, \ie when the $v_\a$ lie in region I.

First let us change variables from $x$ to $t=3\b\sqrt{v_A\,v_B\,v_C}\,x$, a slightly different rescaling than that used in \eqref{RSELE} which will be convenient in the work that follows.  With this change the ELE become
\be \label{elelvform}
\begin{array}{l}
\displaystyle \dot{\p}_\a=\p_\a\left(u_{\a+1}\,\p_{\a+2}-u_{\a+2}\,\p_{\a+1}\right),
\end{array}
\ee
where $u_\a=v_\a/\sqrt{v_A\,v_B\,v_C}$; the $u_\a$ satisfy
\be
u_A+u_B+u_C=u_Au_Bu_C.
\ee
After eliminating $\p_C$ via \eqref{sum} and $u_C$ via $u_C=(u_A+u_B)/(u_A\,u_B-1)$, we obtain the equations
\be
\begin{array}{l}
\displaystyle \dot{\p}_A=\p_A\left[u_B\left(1-\p_A\right)-\frac{u_A\left(1+u_B^2\right)}{u_A\,u_B-1}\,\p_B\right],\\
\displaystyle \dot{\p}_B=-\p_B\left[u_A\left(1-\p_B\right)-\frac{u_B\left(1+u_A^2\right)}{u_A\,u_B-1}\,\p_A\right].
\end{array}
\ee
In this form $u_A$ and $u_B$ are arbitrary positive constants satisfying $u_A\,u_B>1$, and the stationary point is
\be \label{stationarypt}
v_A=\frac{u_A\,u_B-1}{u_B\left(u_A+u_B\right)},\qquad v_B=\frac{u_A\,u_B-1}{u_A\left(u_A+u_B\right)}.
\ee

We next shift the origin in the phase plane to the stationary point \eqref{stationarypt} and make a linear change of variables:
\be
\left[\begin{matrix} x \\ y \end{matrix}\right] = T \left[\begin{matrix} \p_A-v_A \\ \p_B-v_B \end{matrix}\right],
\qquad T = \frac{1}{u_A\,u_B-1}\left[\begin{matrix} u_B\left(1+u_A^2\right) & u_A\left(u_A\,u_B-1\right) \\
                                                   0                       & u_A\left(u_A+u_B\right)    \\
                                    \end{matrix}\right].
\ee
In these new variables the equations become
\be \label{generalizedlv}
\begin{array}{l}
\displaystyle \dot{x} = -y-bx^2-cxy+by^2, \\
\displaystyle \dot{y} = x+xy,
\end{array}
\ee
where
\be \label{lvvars}
b=\frac{u_A\,u_B-1}{1+u_A^2},\qquad c=\frac{2u_A+u_B-u_A^2\,u_B}{1+u_A^2}.
\ee
This is the canonical form for the generalized Lotka-Volterra family of centers presented in equation $(1)$ of \cite{V}.

Given parameters $b$ and $c$ with $b>0$ the equations \eqref{lvvars} can be solved to find
\be
u_A=\frac{-c+\sqrt{c^2+4b(b+1)}}{2b},\qquad u_B=\frac{4b(b+1)}{-c+\sqrt{c^2+4b(b+1)}}-c.
\ee
Clearly $u_A>0$ and one finds easily that $u_A\,u_B>1$, so $u_B>0$. Thus the set of systems \eqref{elelvform} with $u_A$, $u_B$, $u_C>0$ corresponds exactly with the family of all generalized Lotka-Volterra systems \eqref{generalizedlv} with $b>0$.

\section{Restriction on the type of solutions for $r_\a\neq v_\a$, with $v_\a$ in region I} \label{excludedn}

\begin{figure}
\centerline{\includegraphics[scale=.5]{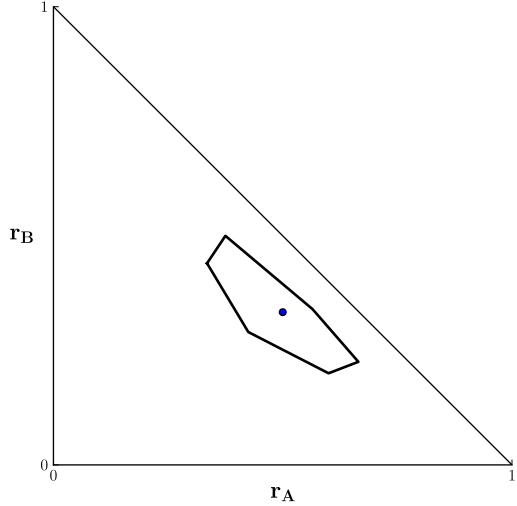}}
\caption{Restriction on the existence of solutions of type $n=2$ in the $r_\a\neq v_\a$ case for $v_A=1/2$, $v_B=1/3$, $v_C=1/6$, using \eqref{rnalpha}. For values of $(r_A,\,r_B)$ outside the bounded region, only type 1 solutions are possible. For $n>2$, the bounded region in which type $n$ solutions are possible would be smaller.}
\label{fig:excludedn}
\end{figure}

A naive estimate of the cutoff $n_{max}$, as described in Section~\ref{solutions}, may be made in the following way. Let us begin with a set of type 1 profiles, with average densities $\tilde{r}_\a$. These profiles will be a portion of the type 1 solution for the $r_\a=v_\a$ case, stretched such that less than one full period of the $\p_\a(x)$ fits inside the interval in $x\in[0,1]$. We will define the length of the profiles $l$, $0<l<1$, to be the fraction of one full period of the $\p_\a(x)$ inside the interval. From these type 1 profiles we may make a set of type $n$ profiles that satisfy the same boundary conditions as the original by rescaling $x$ for the $\p_\a(x)$, such that the original type 1 profiles plus $n-1$ full periods now appear in the interval. Then the average value of the densities for the new type $n$ profiles, $r_{n,\a}$, will be given by
\be \label{rnalpha}
r_{n,\a}=\frac{\tilde{r}_\a\,l+n\,v_\a}{l+n}.
\ee
As a type $n$ solution of the ELE for $r_\a\neq v_\a$ will be a profile of this form, if the $l$ and $\tilde{r}_\a$ in \eqref{rnalpha} cannot be chosen such that the $r_{n,\a}$ are equal to the specified $r_\a$, then this implies that a type $n$ solution does not exist for that case. In Figure~\ref{fig:excludedn} we show an example of the restriction imposed by this simple estimate.

\end{document}